\documentclass{svproc}
\usepackage[round,sort&compress]{natbib}
\usepackage{xspace}
\usepackage{graphicx}
\usepackage{booktabs}
\usepackage{tabularx}
\usepackage{makeidx}  
\makeindex
\usepackage[hyphens]{url}

\makeatletter
\g@addto@macro\UrlBreaks{\do\/\do\_}
\makeatother
\frontmatter          
\graphicspath{{paper_figures_final/}{figures/}}

\newcommand{\kgco}{kg CO$_2$e}
\newcommand{\cuca}{CUCA\xspace}
\newcommand{\orcidlink}[1]{\textsuperscript{\,ORCID: #1}}

\usepackage{etoolbox}
\AtBeginEnvironment{table}{\renewcommand{\arraystretch}{0.94}}
\patchcmd{\thebibliography}{\clearpage}{}{}{}

\begin{document}
\mainmatter

\title{Night-Window Batching versus Carbon-Aware Scheduling for Clinical AI GPU Workloads}
\titlerunning{Night-Window vs.\ Carbon-Aware GPU Scheduling}
\toctitle{Night-Window Batching versus Carbon-Aware Scheduling for Clinical AI GPU Workloads}

\author{Nishi Doshi\inst{1}\orcidlink{0009-0007-8148-4834} \and Shrey Shah\inst{1}}
\authorrunning{N.\ Doshi and S.\ Shah}
\tocauthor{Nishi Doshi and Shrey Shah}

\institute{University of Southern California, Los Angeles, USA\\
\email{nishimit@usc.edu, shrey@alumni.usc.edu}}

\maketitle
\thispagestyle{empty}

\begin{abstract}
\begin{sloppypar}
Hospitals run more machine learning on GPUs while the carbon footprint of grid electricity rises and falls through the day. Using a computer simulation, we compare $13$ scheduling rules on mixed GPU hardware, with synthetic patient-style jobs, urgency tiers, and time-of-day carbon traces. We do not study patient outcomes; every percentage we report is a simulator queue number, not a clinical finding. We ask whether running non-urgent jobs overnight is almost as good as a richer rule that mixes urgency and carbon (CUCA at weight 0.45, written \cuca$_{0.45}$). The comparison keeps carbon reduction secondary to clinical priority and deadline compliance, so each policy is judged on both average \kgco{} and missed-deadline behavior. CarbonGreedy and CarbonShift are carbon-first stress tests that demonstrate how poorly wrong vendor presets can disrupt clinical priorities, and are not meant for production. Numbers are averages over many test settings, with wide run-to-run spread and no statistical adjustment, so headline ratios are exploratory. On an eight-GPU baseline, the overnight rule closes about $78\%$ of the carbon gap between urgency-only and \cuca$_{0.45}$ while missing fewer urgent deadlines than either. CarbonShift lets about $46\%$  of the most urgent jobs miss their deadline; this is simulated queueing, not bedside harm. At $48$ jobs per hour, the carbon footprints almost tie, yet the overnight rule still misses fewer urgent deadlines. A geography test, where regions share one daily carbon shape with only timezone shifts, trims under one percentage point of average carbon; a twelve-hour routine window saves a little carbon for \cuca$_{0.45}$ but raises overall missed deadlines. Overnight batching stays competitive on average modelled carbon; carbon-only rules belong only in stress tests.
\end{sloppypar}
\keywords{medical AI, smart hospital, clinical AI operations, carbon-aware computing, GPU scheduling, sustainable healthcare}
\end{abstract}

\begin{sloppypar}
\noindent\textbf{For hospital operations.}
If you run GPU queues for hospital AI under deadlines, this paper is meant to sit next to your capacity planning. It compares a richer carbon-aware system against a simpler ``run the heavy jobs at night'' approach. Everything here is simulated. Treat the takeaways as hypotheses, unless your job mix, deadlines, hardware, and local grid curves look like ours. Section~\ref{sec:discussion} lists what would still need to be measured on site.
\end{sloppypar}

\section{Introduction}
\label{sec:intro}

\begin{sloppypar}
Hospitals run more and more medical AI on GPU clusters for both training and inference \citep{topol2019deep,rajpurkar2017chexnet}. Cloud operators already move batch work in time and across regions to follow lower-carbon electricity \citep{wiesner2021waitawhile,anderson2022treehouse,acun2023carbonexplorer,souza2023ecovisor}. Healthcare sustainability research keeps track of how big the sector's carbon footprint is and how fast net-zero pressure is rising \citep{eckelman2016environmental,lenzen2020footprint,tennison2021healthcare,nhs2022netzero}.
\end{sloppypar}

\begin{sloppypar}
Hospital leadership rarely cares which scheduler wins. The real question is when more complex scheduling is worth its budget line. Net-zero pledges, vendor ``green mode'' presets, and clinical service-level expectations often pull in different directions, and without shared units you cannot compare those tradeoffs \citep{davenport2019potential,reddy2020governance}. A carbon dashboard with no missed-deadline counter can quietly push the risk of late critical jobs onto clinical staff. A simulation cannot decide policy on its own, but it can show how far a simple ``run heavy jobs at night'' rule moves average carbon and deadline risk before paying for more complex scheduling. It can also flag presets that fall apart under stress and should never run in production. The tables below report results in the same units, so finance, facilities, and clinical IT can talk about the same numbers.
\end{sloppypar}

\begin{sloppypar}
GPU queues for hospital AI differ from regular datacenter jobs because many jobs carry hard deadlines and clinical priority tiers. Real-time scheduling theory teaches the basics of deadlines, priorities, and slack \citep{liu1973scheduling,butazzo2011realtime,stankovic1995deadline}, but it does not tell us whether a simple night-window rule is already enough, or whether a richer carbon-aware rule is worth the extra complexity. Work on ML energy and carbon has also argued for moving training and inference to greener times and places \citep{strubell2019energy,schwartz2020green,patterson2021carbon,dodge2022measuring,patterson2022carbonplateau,luccioni2024powerhungry}.
\end{sloppypar}

\begin{sloppypar}
We run a \textbf{side-by-side comparison}: one simulator, thirteen schedulers on a mix of GPUs, synthetic workloads, and carbon-only \emph{stress tests} that show what happens when carbon-only choices break clinical priorities (we are not proposing them as policies for hospital use). Think of it as lining up thirteen queueing rules under the same stopwatch. Three experiment families in Section~\ref{sec:methods} repeat the same patterns. On the full gate grid, the overnight rule closes most of the average carbon gap from urgency-only to \cuca$_{0.45}$, with fewer simulator critical-tier misses. By design, the stress rules miss many deadlines. At 48 jobs per hour, average carbon almost ties while the night window stays safer on critical misses. Geo adds under $1\%$ extra carbon savings, and only when every region shares one daily curve shifted by timezone (a small geography sanity check, not a general claim). A 12-hour routine longshift trims about $0.8\%$ carbon but raises overall missed deadlines. We report patterns that depend on the setting; there is no single winner.
\end{sloppypar}

\paragraph{Contributions.}
\begin{itemize}\setlength{\itemsep}{1pt}\setlength{\parsep}{0pt}
    \item We run thirteen policies against the same workload and cluster model, with identical simulator rules and the same post-run metrics for every run.
    \item In the gate study, averaged over all settings, the overnight rule closes about $78\%$ of the average \kgco{} gap from urgency-only to \cuca$_{0.45}$, while critical-tier misses are about $2.4\times$ lower. \textbf{CarbonGreedy} and \textbf{CarbonShift} are deliberately unsafe carbon-only stress runs (queue numbers only); they show what bad presets look like, not policies you would deploy.
    \item At 48 jobs per hour, the overnight rule matches or slightly beats \cuca$_{0.45}$ on average carbon, with fewer critical misses. In the geo battery (one daily curve per region, shifted only by timezone), geo gives less than $1\%$ extra average \kgco{} over the single-region night baseline we compare against. This is not a claim about real tariffs or data-sovereignty tradeoffs.
    \item A 720-minute routine longshift trims about $0.8\%$ average \kgco{} but raises overall simulator misses from $3.75\%$ to $5.08\%$ on the geo+longshift grid. These patterns depend on our settings; they are not a single ranking for every hospital.
\end{itemize}

\section{Related Work}
\label{sec:related}

\begin{sloppypar}
We study \emph{batch} GPU scheduling when grid carbon intensity changes through the day and jobs have deadlines. The question is whether running deferrable jobs at night is enough, or whether a richer carbon-aware policy is worth the added complexity. The setting is motivated by hospital training and inference queues, but the question is operational, not clinical.
\end{sloppypar}

\begin{sloppypar}
Most carbon-aware computing work targets datacenters and cloud regions, moving work in time or across regions to cut emissions, usually without clinical-style urgency \citep{wiesner2021waitawhile,anderson2022treehouse,acun2023carbonexplorer,souza2023ecovisor}. Real-time scheduling gives the standard language of feasibility, slack, and urgency when deadlines matter \citep{liu1973scheduling,butazzo2011realtime,stankovic1995deadline}. In medical imaging, deep models already support screening tasks such as diabetic retinopathy and chest radiography \citep{gulshan2016development,doshi2020diabetic,rajpurkar2017chexnet}. Other work estimates healthcare's carbon footprint \citep{eckelman2016environmental,lenzen2020footprint,tennison2021healthcare,nhs2022netzero}, while ML systems papers measure training and inference emissions and argue for greener times or regions \citep{strubell2019energy,schwartz2020green,patterson2021carbon,dodge2022measuring,patterson2022carbonplateau,luccioni2024powerhungry}. We combine these threads for one synthetic hospital GPU workload, one metric stack, and thirteen policies.
\end{sloppypar}

\section{Model, Policies, and Experiments}
\label{sec:methods}

\paragraph{Jobs, GPUs, and carbon accounting.}
Each job has an arrival time, a runtime, an urgency label $u_j\in\{1,2,3\}$ (routine, urgent, \emph{critical}), and a hard deadline. GPUs differ in speed, power draw, and the local grid carbon intensity $g_r(t)$, which varies through the day. Per-job \kgco{} is the energy used while the job runs, multiplied by the average $g_r(t)$ over that interval. This is operational carbon at the GPU's location; Scope~3 (the carbon from making the GPU and its full lifecycle) is out of scope.

\paragraph{Metrics and simulator.}
A \emph{critical deadline miss} is the share of finished critical jobs ($u_j{=}3$) that finish after their deadline. This is a queue number from the simulator, not a measure of patient harm. We report the average total \kgco, the p95 turnaround for critical jobs, the critical and overall deadline-miss rates, and weighted tardiness. All policies are simple rules running in the same event-by-event simulator. After every run we compute the same metrics from finished jobs, so scheduler names stay directly comparable.

\begin{sloppypar}
\textbf{What one run looks like.}
Each configuration draws 2000 jobs with a fixed arrival curve and carbon trace, then changes only the random seed. We average the metrics across all combinations of arrival rate, critical fraction, and scenario, and read the overall row first, then the breakdowns.
\end{sloppypar}
\begin{sloppypar}
The roster groups into five families. \textbf{FIFO} and \textbf{SJF} are arrival-only baselines: whoever arrived first, or whoever looks shortest. \textbf{EDF} and \textbf{UrgencyOnly} sort by deadline or urgency tier (the same on the gate grid). \textbf{CarbonGreedy} and \textbf{CarbonShift} are carbon-only stress tests. They show how badly carbon-only tuning can break clinical ordering and are not schedulers we propose. \textbf{NightWindowDefer}, \textbf{MiddayWindowDefer}, and \textbf{SingleRegionNightDefer\_*} push deferrable work into low-carbon windows; \textbf{SlackAwareCarbon} mixes slack checks with carbon placement. \textbf{\cuca$_{\alpha}$} blends urgency and carbon at weight $\alpha\in\{0.90,0.75,0.60,0.45\}$. We also run \cuca$_{0.45}$\_longshift with a 720-minute routine horizon. \cuca{} ranks by urgency and deadline, preserves critical slack, searches for deadline-safe deferral, then runs immediately if no safe deferral exists.
\end{sloppypar}

\paragraph{Batteries.}
\begin{sloppypar}
\textbf{(1) Gate:} \path{heterogeneous_8gpu}, 2000 jobs/run, seeds $\{1,\ldots,5\}$, arrivals $\{24,36,48\}$/h, critical fractions $\{0.08,0.12,0.18\}$, scenarios \path{normal} / \path{volatile} / \path{renewable_midday}. \textbf{(2) Geo:} \path{multi_region_global}, five regions, one diurnal template phase-shifted by timezone (no extra inter-region noise). This is a routing sensitivity check, not a full migration model. \textbf{(3) Longshift:} battery~(2) plus \path{CUCA_a0.45_longshift}, \path{CUCA_a0.75_longshift}.
\end{sloppypar}

\begin{sloppypar}
All three batteries reuse the same workload generator and reporting scripts, and the tables and figures regenerate from the checked-in CSV files. With many policy-by-condition pairs, we treat side-by-side gaps as descriptive observations, not as final rankings on unseen workloads.
\end{sloppypar}

\paragraph{What this study cannot prove.}
\begin{enumerate}\setlength{\itemsep}{0pt}\setlength{\parsep}{0pt}\setlength{\topsep}{2pt}
  \item No patient outcomes; \emph{critical-tier miss} is a queue number from the simulator.
  \item Synthetic arrivals only; real hospital traces probably have busier hours and downtimes we do not model.
  \item Carbon is operational intensity at the GPU's location, averaged per job. Marginal carbon and Scope~3 (the carbon from making the GPU and its full lifecycle) are out of scope.
  \item The geo battery shares one daily carbon \emph{shape} across regions; it is a sensitivity check on timezone shifts, not a test on real curves.
  \item We do not model energy or latency overheads from job migration, regional workload redistribution, data transfer, remote storage access, or orchestration.
  \item Five seeds, descriptive averages, no statistical adjustment; headline ratios are exploratory.
\end{enumerate}

\section{Results}
\label{sec:results}

Results follow the three batteries. Tables show the full settings; the text highlights gaps between policies that hold up across them.

\begin{sloppypar}
\textbf{Reading order.}
Table~\ref{tab:gate_full} and Figure~\ref{fig:gate} carry the deadline-versus-carbon picture, and the arrival-rate tables follow for readers who care most about utilization. Read the geo numbers alongside the shared-curve design from Section~\ref{sec:methods}; otherwise small gaps look like a bug, not a modelling choice.
\end{sloppypar}

\subsection{Section~1 gate: full scheduler roster}
\label{sec:res:gate}

The gate roster covers FIFO/SJF, EDF/UrgencyOnly, stress policies, night and midday windows, slack-aware rules, and four \cuca{} weights. Table~\ref{tab:gate_full} averages over all settings. Among non-stress policies, FIFO has the worst simulator critical-tier misses. EDF, UrgencyOnly, and \cuca$_{0.90}$ coincide on the overall averages. SJF is gentler on tail latency than FIFO but still ignores clinical carbon trade-offs. We keep both because unattended clusters often drift toward one of them without anyone naming a ``policy.''

\begin{table}[t]
\centering
\small
\caption{Gate study: mean metrics over the full grid.}
\label{tab:gate_full}
\scriptsize
\setlength{\tabcolsep}{3pt}
\resizebox{\linewidth}{!}{
\begin{tabular}{lrrrr}
\toprule
Scheduler & Carbon (\kgco) & Crit.\ p95 (min) & Crit.\ miss (\%) & Overall miss (\%) \\
\midrule
FIFO & 52.04 & 23.45 & 16.840 & 2.819 \\
SJF & 52.02 & 9.08 & 0.541 & 0.204 \\
EDF & 52.03 & 8.46 & 0.154 & 0.017 \\
UrgencyOnly & 52.03 & 8.46 & 0.154 & 0.017 \\
CarbonGreedy & 44.09 & 24.21 & 17.213 & 2.894 \\
CarbonShift & 44.06 & 165.70 & 45.937 & 18.967 \\
NightWindowDefer & 46.35 & 8.61 & 0.080 & 0.186 \\
MiddayWindowDefer & 47.80 & 8.66 & 0.217 & 2.068 \\
SlackAwareCarbon & 44.69 & 8.81 & 0.152 & 0.469 \\
CUCA\_a0.90 & 52.03 & 8.46 & 0.154 & 0.017 \\
CUCA\_a0.75 & 50.10 & 9.02 & 0.177 & 0.046 \\
CUCA\_a0.60 & 49.17 & 9.04 & 0.167 & 1.146 \\
CUCA\_a0.45 & 44.71 & 8.94 & 0.190 & 1.694 \\
\bottomrule
\end{tabular}
}
\end{table}

\begin{figure}[t]
\centering
\includegraphics[width=0.238\linewidth,trim=0 169.8bp 200.2bp 0,clip]{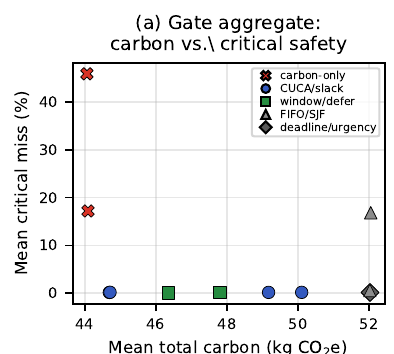}%
\hfill
\includegraphics[width=0.238\linewidth,trim=200.2bp 169.8bp 0 0,clip]{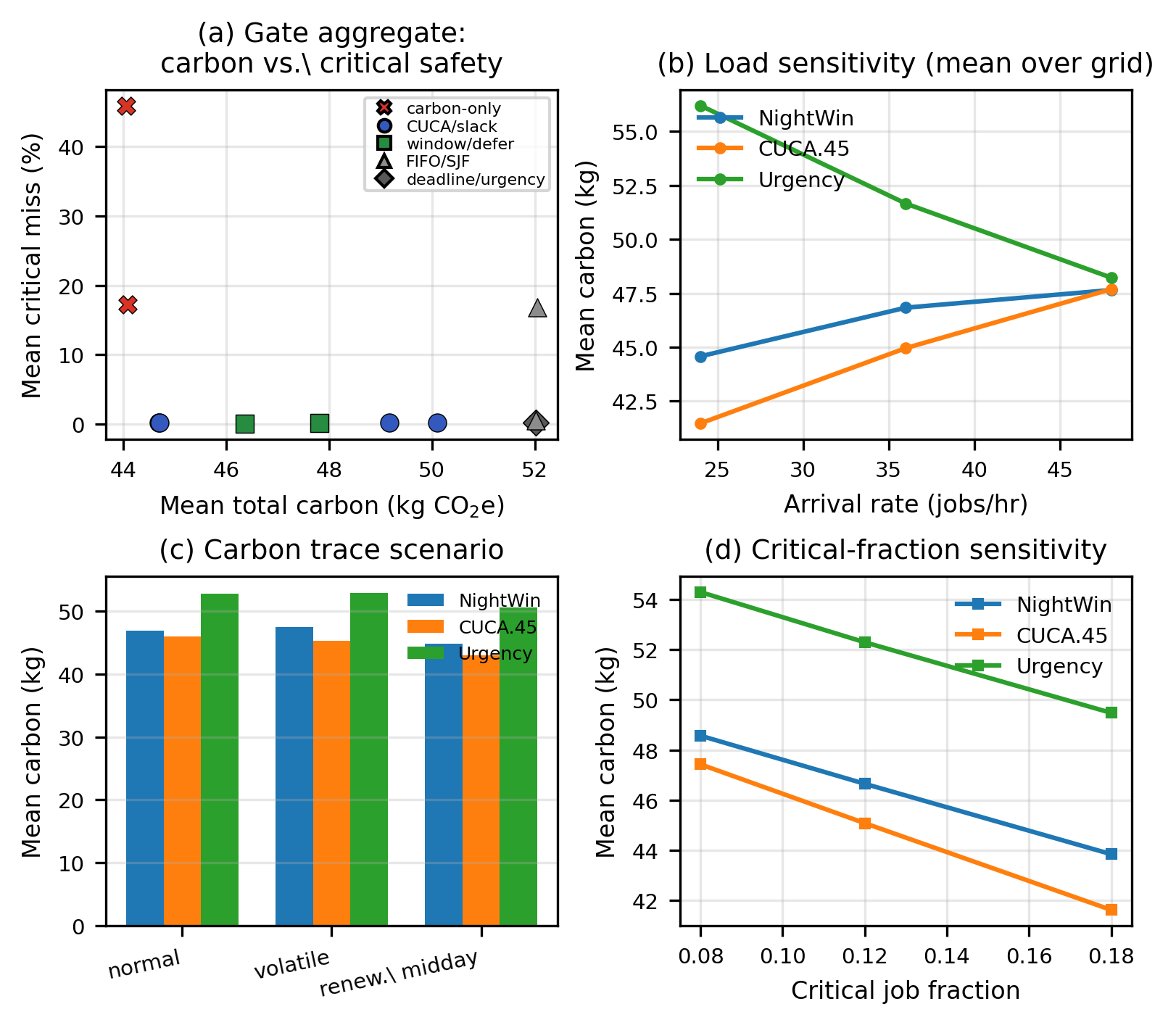}%
\hfill
\includegraphics[width=0.238\linewidth,trim=0 0 200.2bp 169.8bp,clip]{fig_gate_composite}%
\hfill
\includegraphics[width=0.238\linewidth,trim=200.2bp 0 0 169.8bp,clip]{fig_gate_composite}
\caption{\textbf{Gate battery.} (a)~Mean \kgco{} vs.\ simulator critical-tier miss rate. (b)--(d)~Mean \kgco{} for UrgencyOnly, NightWindowDefer, and \cuca$_{0.45}$ by arrival rate, carbon scenario, and critical-job fraction; other factors are averaged.}
\label{fig:gate}
\end{figure}

Figure~\ref{fig:gate}: (a)~grid-mean \kgco{} vs.\ \emph{simulator} critical-tier miss (\%); (b)--(d)~mean \kgco{} vs.\ arrival rate, carbon scenario, and critical-fraction for UrgencyOnly, NightWindowDefer, and \cuca$_{0.45}$ (other factors averaged within each panel; see caption).

\subsection{\cuca{} $\alpha$ ablation on the gate grid}
\label{sec:res:cuca}

\begin{figure}[t]
\centering
\begin{minipage}[t]{0.49\linewidth}
\centering
\includegraphics[width=\linewidth,height=0.13\textheight,keepaspectratio]{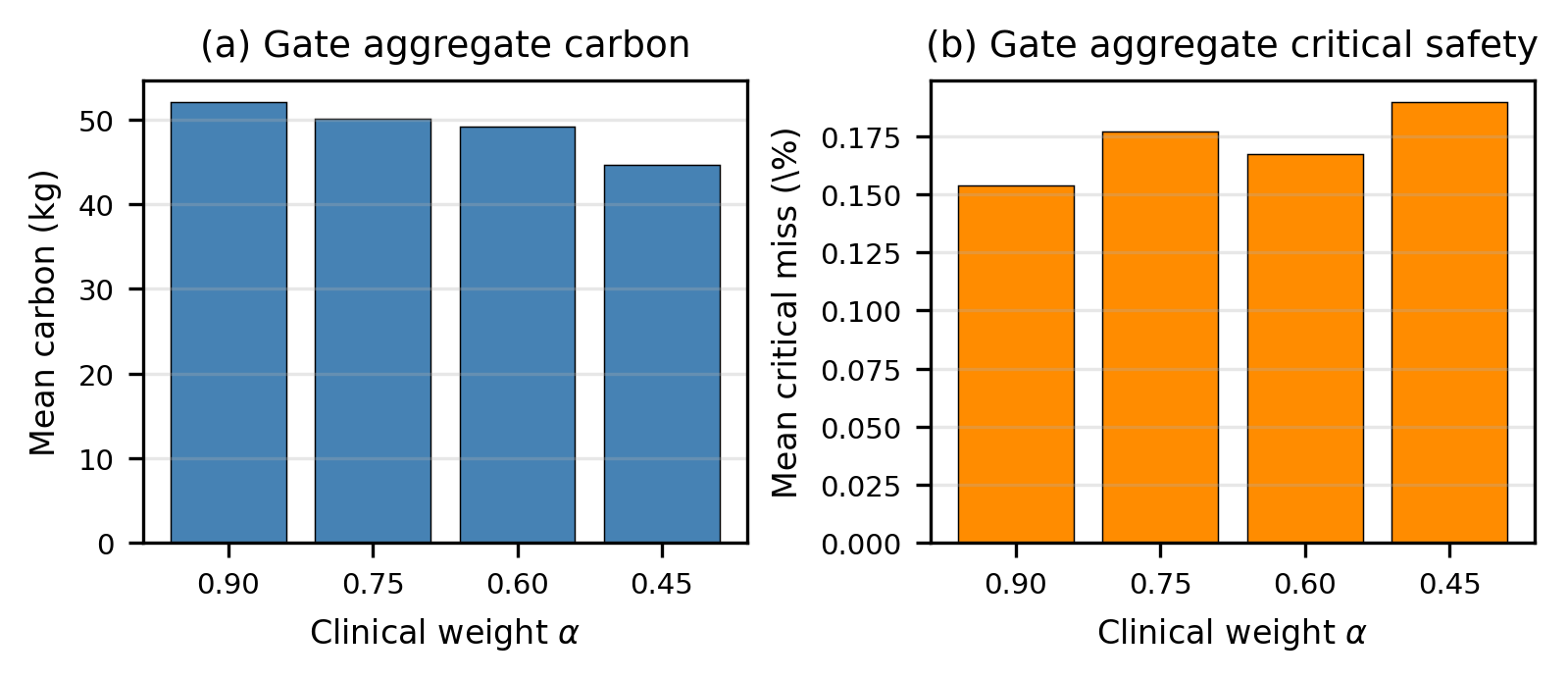}
\caption{\textbf{\cuca{} sweep.} Gate-grid means: (a)~\kgco{} vs.\ $\alpha$; (b)~\emph{simulator} critical-tier miss rate vs.\ $\alpha$.}
\label{fig:cuca_alpha}
\end{minipage}\hfill
\begin{minipage}[t]{0.49\linewidth}
\centering
\includegraphics[width=\linewidth,height=0.13\textheight,keepaspectratio]{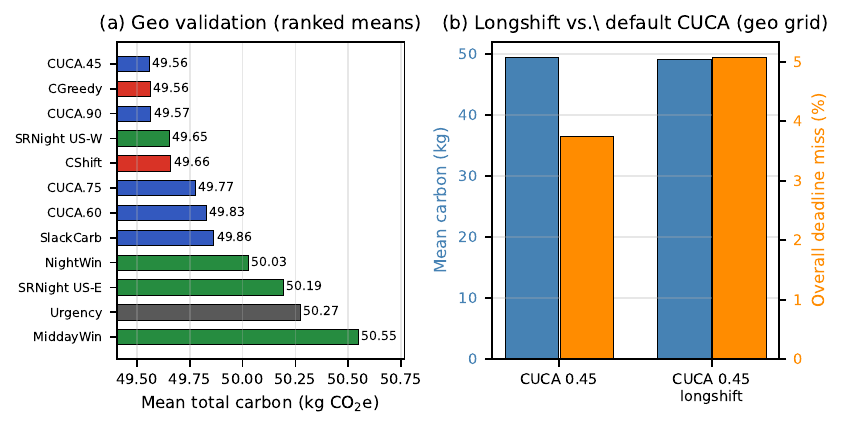}
\caption{\textbf{Geo and longshift.} Shared daily carbon shape with timezone shifts only. (a)~Geo mean \kgco{} ranking; (b)~\cuca$_{0.45}$ default vs.\ 720-minute longshift.}
\label{fig:geo}
\end{minipage}
\end{figure}

Figure~\ref{fig:cuca_alpha} sweeps $\alpha$ on the gate grid: panel~(a) shows average total \kgco{} versus $\alpha$, and panel~(b) shows the average simulator critical-tier miss rate. Average carbon falls as $\alpha$ drops; critical misses stay small but rise toward the more aggressive end of the sweep. This is the safety-versus-carbon knob an operator tunes when picking the weight at which to run \cuca{}.

\subsection{Stratification by arrival rate (gate)}
\label{sec:res:strat}

Tables~\ref{tab:gate_arr_c} and \ref{tab:gate_arr_m} average across seeds with the scenario and critical fraction fixed per run. NightWindowDefer beats UrgencyOnly on average carbon at every listed arrival rate. \cuca$_{0.45}$ sits between them. \textbf{CarbonShift} is a carbon-only \emph{stress test}, not a clinical policy: it keeps very low average \kgco{} and very high simulator critical-tier miss rates at each arrival rate (queue fractions, not patient harm; see Table~\ref{tab:gate_arr_m}).

\begin{table}[t]
\centering
\small
\begin{minipage}[t]{0.49\linewidth}
\centering
\caption{Gate study: mean total carbon (\kgco) by arrival rate (jobs/hour).}
\label{tab:gate_arr_c}
\scriptsize
\setlength{\tabcolsep}{2pt}
\resizebox{\linewidth}{!}{
\begin{tabular}{lrrr}
\toprule
Scheduler & 24 j/h & 36 j/h & 48 j/h \\
\midrule
UrgencyOnly & 56.20 & 51.66 & 48.22 \\
NightWindowDefer & 44.58 & 46.83 & 47.65 \\
MiddayWindowDefer & 44.97 & 48.80 & 49.62 \\
SlackAwareCarbon & 41.23 & 45.19 & 47.64 \\
CUCA\_a0.45 & 41.47 & 44.96 & 47.69 \\
CarbonShift & 40.19 & 44.60 & 47.37 \\
\midrule
\multicolumn{4}{l}{\scriptsize Means over seeds, critical fractions, scenarios.} \\
\bottomrule
\end{tabular}
}
\end{minipage}\hfill
\begin{minipage}[t]{0.49\linewidth}
\centering
\caption{Gate study: mean critical deadline miss rate (\%) by arrival rate.}
\label{tab:gate_arr_m}
\scriptsize
\setlength{\tabcolsep}{2pt}
\resizebox{\linewidth}{!}{
\begin{tabular}{lrrr}
\toprule
Scheduler & 24 j/h & 36 j/h & 48 j/h \\
\midrule
UrgencyOnly & 0.000 & 0.044 & 0.418 \\
NightWindowDefer & 0.047 & 0.101 & 0.093 \\
MiddayWindowDefer & 0.091 & 0.179 & 0.381 \\
SlackAwareCarbon & 0.073 & 0.179 & 0.205 \\
CUCA\_a0.45 & 0.081 & 0.225 & 0.262 \\
CarbonShift & 20.655 & 48.213 & 68.943 \\
\bottomrule
\end{tabular}
}
\end{minipage}
\end{table}

\subsection{Geo validation: aggregate roster}
\label{sec:res:geo}

\begin{sloppypar}
Table~\ref{tab:geo_full} mirrors the gate layout on \path{multi_region_global}. \path{SingleRegionNightDefer_us_west} is the strongest single-region night anchor we compare. \cuca$_{0.45}$ is only marginally lower on average \kgco{} ($\approx\!0.19\%$). That is the expected result when every region sees the \emph{same} daily carbon curve and differs only by the clock. With identical curves, there is little real room for cross-region routing to save carbon, before counting any migration overhead.
\end{sloppypar}

\begin{sloppypar}
Figure~\ref{fig:geo}: (a)~ranked geo-grid mean \kgco{} (truncated axis, labels in kg; lowest at top); (b)~default \cuca$_{0.45}$ vs.\ 720-minute longshift (Table~\ref{tab:longshift}).
\end{sloppypar}

\begin{table}[t]
\centering
\small
\caption{Geo validation: mean metrics over the geo factorial grid.}
\label{tab:geo_full}
\scriptsize
\setlength{\tabcolsep}{3pt}
\resizebox{\linewidth}{!}{
\begin{tabular}{lrrrr}
\toprule
Scheduler & Carbon (\kgco) & Crit.\ p95 (min) & Crit.\ miss (\%) & Overall miss (\%) \\
\midrule
FIFO & 50.26 & 37.02 & 23.744 & 5.387 \\
SJF & 50.28 & 9.50 & 0.619 & 0.430 \\
EDF & 50.27 & 8.71 & 0.134 & 0.018 \\
UrgencyOnly & 50.27 & 8.71 & 0.134 & 0.018 \\
CarbonGreedy & 49.56 & 37.02 & 23.744 & 5.387 \\
CarbonShift & 49.66 & 123.35 & 37.525 & 13.212 \\
NightWindowDefer & 50.03 & 8.98 & 0.124 & 1.264 \\
MiddayWindowDefer & 50.55 & 8.98 & 0.133 & 5.668 \\
SlackAwareCarbon & 49.86 & 8.86 & 0.140 & 1.208 \\
CUCA\_a0.90 & 49.57 & 8.71 & 0.134 & 0.018 \\
CUCA\_a0.75 & 49.77 & 8.74 & 0.107 & 0.013 \\
CUCA\_a0.60 & 49.83 & 8.81 & 0.092 & 0.604 \\
CUCA\_a0.45 & 49.56 & 8.81 & 0.104 & 3.747 \\
SingleRegionNightDefer\_us\_west & 49.65 & 8.99 & 0.145 & 3.666 \\
SingleRegionNightDefer\_us\_east & 50.19 & 9.08 & 0.170 & 1.473 \\
\bottomrule
\end{tabular}
}
\end{table}

\subsection{Longshift on the geo grid}
\label{sec:res:longshift}

\begin{sloppypar}
Table~\ref{tab:longshift} lists the longshift roster. The numeric headlines below match the abstract (all are simulator averages; see the table for full rows). On the gate grid, going from urgency-only to \cuca$_{0.45}$, the overnight rule recovers about $78\%$ of the average \kgco{} saving with about $2.4\times$ fewer simulator critical-tier misses. At 48 jobs per hour, average carbon almost ties at $47.65$ vs.\ $47.69$~\kgco. On the shared-curve geo battery, the geo single-region rule vs.\ \cuca$_{0.45}$ is $49.65$ vs.\ $49.56$~\kgco. The longshift moves average \kgco{} from $49.556$ to $49.150$, while overall simulator misses creep from $3.75\%$ to $5.08\%$.
\end{sloppypar}
\begin{sloppypar}
\cuca$_{0.75}$\_longshift sits between default \cuca$_{0.75}$ and \cuca$_{0.45}$ on carbon with a smaller miss bump than the $0.45$ longshift row.
\end{sloppypar}

\begin{table}[t]
\centering
\small
\caption{Geo + longshift battery: selected policies.}
\label{tab:longshift}
\scriptsize
\setlength{\tabcolsep}{3pt}
\resizebox{\linewidth}{!}{
\begin{tabular}{lrrrr}
\toprule
Scheduler & Carbon (\kgco) & Crit.\ p95 (min) & Crit.\ miss (\%) & Overall miss (\%) \\
\midrule
UrgencyOnly & 50.27 & 8.71 & 0.134 & 0.018 \\
NightWindowDefer & 50.03 & 8.98 & 0.124 & 1.264 \\
SingleRegionNightDefer\_us\_west & 49.65 & 8.99 & 0.145 & 3.666 \\
CUCA\_a0.75 & 49.77 & 8.74 & 0.107 & 0.013 \\
CUCA\_a0.45 & 49.56 & 8.81 & 0.104 & 3.747 \\
CUCA\_a0.75\_longshift & 49.48 & 8.92 & 0.127 & 1.374 \\
CUCA\_a0.45\_longshift & 49.15 & 8.65 & 0.143 & 5.081 \\
\bottomrule
\end{tabular}
}
\end{table}
\vspace{-0.6\baselineskip}

\section{Discussion and Conclusion}
\label{sec:discussion}

\begin{sloppypar}
\textbf{Takeaways.}
Carbon is a secondary goal kept behind deadlines and priority tiers. On the gate grid, the overnight rule keeps most of \cuca$_{0.45}$'s average \kgco{} benefit with fewer simulator critical-tier misses, and at 48 jobs per hour average \kgco{} is essentially tied while the night window is still safer on critical misses. \textbf{CarbonGreedy} and \textbf{CarbonShift} show what bad carbon-only choices look like (Table~\ref{tab:gate_full}); they are queue experiments, not patient outcomes, and should not run in production. Geo gains stay small under our setting where every region shares one daily curve shifted by timezone (Section~\ref{sec:methods}); treat that as a sensitivity check, not a proof that geography never helps. The 12-hour routine longshift saves little \kgco{} ($\approx\!0.8\%$) at the cost of more overall simulator misses. ``Critical'' here means simulator tier $u{=}3$ queueing only \citep{kaissis2020federated,dean2013tail}. Prefer overnight buffers before more complex scheduling, and validate geo on site-specific curves before any procurement decision. Hospitals should also account for migration energy, network transfer, data locality, privacy constraints, and inter-region latency before adopting regional routing. Headline ratios are averages with wide variation (e.g.\ NightWindowDefer vs.\ \cuca$_{0.45}$ std.\ $2.77$ vs.\ $3.88$~\kgco); the comparisons are descriptive and not adjusted for multiple comparisons.
\end{sloppypar}

\begin{sloppypar}
\textbf{Procurement and governance.}
Three actions follow from these numbers. First, acceptance tests for any vendor ``green mode'' should run carbon-only and priority-aware modes side by side. A jump in simulator critical-tier misses is a warning sign about how the system is configured, not a real saving \citep{reddy2020governance}. Second, board-level dashboards benefit from a paired missed-deadline counter, so one kilogram headline cannot quietly hide late critical jobs. Third, widening routine deferral is a change-control decision. Our longshift rows move overall miss rates from $3.75\%$ to $5.08\%$ in exchange for a $0.8\%$ carbon trim, so any longer window needs sign-off from clinical staff.
\end{sloppypar}

\noindent\textbf{Operator playbook.}\par
\begingroup
\sloppy
\scriptsize
\setlength{\tabcolsep}{3pt}%
\renewcommand{\arraystretch}{1.2}%
\noindent
\begin{tabularx}{\linewidth}{@{}>{\raggedright\arraybackslash}X >{\raggedright\arraybackslash}X >{\raggedright\arraybackslash}X@{}}
\toprule
\textbf{If your site resembles\ldots} & \textbf{A reasonable first step} & \textbf{When to explore \cuca{} or geo routing} \\
\midrule
Our gate-style mix: heterogeneous GPUs, smooth diurnal grid curves, and deadline-heavy clinical-style priorities in simulation. &
Document a night-window plus explicit deadline buffers before you buy heavier orchestration. &
Consider \cuca{} weighting only if traces show slack you still need to tune; at the highest loads we test, mean carbon nearly ties simpler rules anyway. \\
\midrule
Regional carbon traces behave like one shared daily shape with only timezone shifts (our geo battery). &
Anchor operations on a single-region night policy you trust. &
Revisit geo shifting after local tariffs show \emph{different} curve shapes or marginal carbon, not only clock offsets. \\
\midrule
You are procuring scheduler software or approving a vendor ``green mode.'' &
Require acceptance tests that include carbon-only stress runs side by side with clinical-priority modes. &
If simulator critical-tier misses explode under carbon-first settings, treat that as a configuration hazard, not a score to chase. \\
\midrule
You are tempted to widen routine deferral windows (our longshift story). &
Treat long horizons as a change-control item tied to tardiness risk. &
Expect overall miss rates to creep when slack widens; pair any long window with explicit clinical sign-off on acceptable delay. \\
\bottomrule
\end{tabularx}%
\par\endgroup

\begin{sloppypar}
\textbf{Conclusion.}
Under our synthetic workloads and daily carbon curves, simple overnight batching is competitive with more complex carbon-aware rules on average carbon, and often improves critical-tier safety. Carbon-only objectives fail badly as stress tests. Geo and longshift give small or costly improvements in the settings we built. The natural next step is the same comparison on real hospital workloads and grid data, with regional diversity beyond a single shared daily curve and explicit migration costs. Before any procurement decision rests on a simulation alone, replay the same policy roster on de-identified arrivals, marginal carbon from tariffs or live grid APIs, and the actual GPU mix.
\end{sloppypar}

\section*{Acknowledgments}
\begin{sloppypar}
The authors thank colleagues and reviewers for comments that improved the comparative framing of this work. AI-assisted programming and editing tools were used during code development, preprocessing, and drafting. The authors reviewed and edited the manuscript and remain responsible for the published content.
\end{sloppypar}

\section*{References}
\begingroup
\footnotesize
\setlength{\bibsep}{0.05ex plus 0.05ex minus 0.05ex}
    \makeatletter
    \renewcommand{\section}[2]{} 
    \renewcommand{\clearpage}{}
    \def\@makeschapterhead#1{} 
    \makeatother

    \bibliographystyle{spbasic}
    \bibliography{references}
\endgroup

\end{document}